\documentclass[aps,prmaterials,reprint,superscriptaddress,floatfix]{revtex4-2}

\usepackage{amssymb}
\usepackage{amsmath}
\usepackage{graphicx}
\usepackage{dcolumn}
\usepackage{bm}

\usepackage{textcomp,gensymb}
\usepackage{balance}
\usepackage{xcolor}

\begin{document}


\title{Growth and Characterization of $\beta$-Mn Structured CoZn Thin Films}

\author{M.~Dearg}
\email[Email:~]{malcolm.dearg@york.ac.uk}
\altaffiliation{Present address: School of Physics, Engineering and Technology, University of York, York YO10 5DD, United Kingdom}
\affiliation{School of Physics and Astronomy, University of Leeds, Leeds LS2 9JT, United Kingdom}

\author{G.~Burnell}
\affiliation{School of Physics and Astronomy, University of Leeds, Leeds LS2 9JT, United Kingdom}

\author{S.~Langridge}
\affiliation{ISIS Neutron and Muon Source, STFC Rutherford Appleton Laboratory, Didcot, Oxon. OX11 0QX, United Kingdom}

\author{C.~H.~Marrows}
\email[Email:~]{c.h.marrows@leeds.ac.uk}
\affiliation{School of Physics and Astronomy, University of Leeds, Leeds LS2 9JT, United Kingdom}


\date{\today}

\begin{abstract}
Thin films of polycrystalline $\beta$-Mn structure CoZn have been grown on thermally oxidized Si substrates by co-sputtering from elemental targets followed by annealing. A range of films grown with variable Co deposition power and fixed Zn deposition power were produced, so as to vary the proportions of the two elements reaching the substrate, which were annealed post-growth. Whilst all films exhibited a (221) $\beta$-Mn structure CoZn texture in X-ray diffraction, transmission electron microscopy showed that the composition with the highest integrated intensity for that Bragg peak contained large vacancies and was covered by a thick ZnO cap owing to being Co-deficient overall. CoZn films deposited at ratios tuned to give the optimal volume fraction of $\beta$-Mn were continuous, with crystallites up to 200~nm in size, with a much thinner ZnO cap layer. Magnetic measurements show that such optimal CoZn films have a Curie temperature $T_\mathrm{C} \sim 420$~K and saturation magnetization of 120~emu/cm$^3$, properties close to those reported for bulk crystals. The $\beta$-Mn structure is chiral (P4$_{1}$32/P4$_{3}$32 space group) and is known to give rise to a Dzyaloshinkii-Moriya interaction (DMI) that stabilizes room-temperature skyrmions in the bulk. Our thin films are thus a potential materials platform, compatible with planar processing technology, for magnetic skyrmions arising from a bulk DMI.
\end{abstract}


\maketitle 

\section{Introduction\label{sec:intro}}

Magnetic skyrmions, topologically non-trivial magnetization textures with particle-like properties, hold potential for data storage due to their topological stability, nanometric size, and low driving current densities \cite{Fert2013,Finocchio2016,Marrows2021}. Long-predicted as arising in magnets with chiral interactions \cite{Bogdanov1989}, they were first observed in MnSi \cite{Muhlbauer2009c}, which possesses the B20 lattice structure. This chiral lattice gives rise to a bulk Dzyaloshinskii-Moriya interaction (DMI) that stabilizes the skyrmions. Skyrmions were subsequently observed in other B20 compounds such as Fe$_{1-x}$Co$_x$Si \cite{Yu2010}, FeGe \cite{Yu2011}, and Cu$_2$OSeO$_3$ \cite{Seki2012}. Nevertheless, these widely-studied B20 skyrmion materials present a bottleneck to applications, owing to the fact that they only exhibit magnetic order below room-temperature \cite{Kanazawa2017b}.

\begin{figure}
    \includegraphics[width=8.6cm]{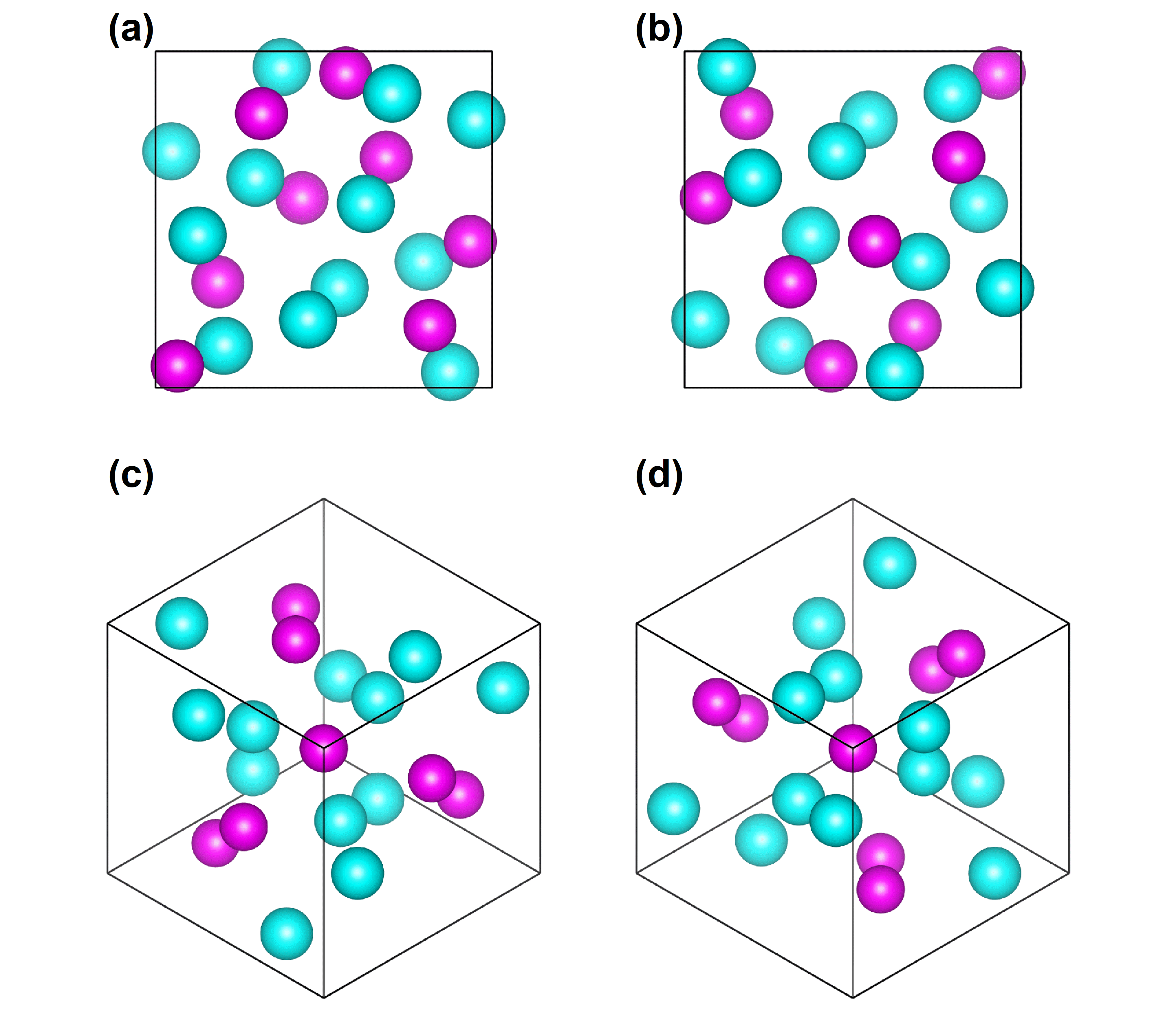}
	\caption{Crystallographic visualization of the $\beta$-Mn structure P4$_{1}$32 and P4$_{3}$32 space groups, for a 293~K  CoZn alloy of lattice constant 6.33~\AA , and equivalent isotropic displacement [x, y, z] parameters for 8c (magenta) and 12d (cyan) Wyckoff sites of [0.0648, 0.0648, 0.0648]  and [0.125, 0.2030, 0.4530]  respectively \cite{Xie2013c}. (a,c) [100] \& [111] orientations of P4$_{1}$32, respectively. (b,d) [100] \& [111] orientations of P4$_{3}$32, respectively. 
    Visualizations were produced using VESTA \cite{Momma2011}. \label{fig:xtal}}
\end{figure}

By revisiting non-centrosymmetric crystals reported to be ferromagnetic above room temperatures,
a new class of material was discovered to host skyrmions above room-temperature \cite{Tokunaga2015}.
CoZnMn alloys have long been known to host a $\beta$-Mn phase \cite{Buschow1983,Hori2007,Karlsen2009,Xie2013c}. The $\beta$-Mn structure, named for the second allotrope of Mn, refers to the cubic space group numbers 212 and 213, which represent the chiral handedness due to the enantiomorphic nature of point group 432 \cite{Cockcroft1999}. The $\beta$-Mn-type structure forms a primitive cubic of space group P4$_{3}$32 or P4$_{1}$32 depending on handedness (shown in Fig.~\ref{fig:xtal}), which consists of 20 atoms split across the inequivalent crystallographic 8c and 12d sites \cite{Donohue1974}. This structure breaks spatial inversion symmetry, the condition required for a bulk DMI to manifest itself. Such alloys were shown to exhibit DMI-stabilized skyrmions at and above room temperature in bulk \cite{Tokunaga2015}, with a rich variety of properties subsequently discovered \cite{Karube2016a,Karube2017b,Kanazawa2017b,Takagi2017,Morikawa2017a,Karube2018a,Bocarsly2019}.

This makes such $\beta$-Mn structure alloys candidates for developing room temperature skyrmion technologies based on bulk DMI, alongside magnetic multilayer systems that possess interfacial DMI \cite{Chen2015,Jiang2015,Moreau-Luchaire2016,Woo2016,Boulle2016,Soumyanarayanan2017,Zeissler2018}. Thin film materials are needed for compatibility with the planar processing methods used in the microelectronics industry. Whilst the multilayer systems are already in this form, $\beta$-Mn structure CoZn-based alloys have only been studied as bulk crystals to date.

Here we show the development of a co-sputter deposition and post-growth annealing protocol capable of yielding CoZn films with the desired $\beta$-Mn crystal structure. Since the films are grown on the amorphous SiO$_x$ surface of thermally oxidized Si wafer, they are polycrystalline, albeit with large grains and a strong $(221)$ texture. We emphasize the importance of correctly balancing the sputter fluxes of Co and Zn reaching the substrate. The high mobility of Zn means that excess Zn segregates to the surface of the film and oxidizes, forming a ZnO capping layer, leaving behind vacancies for Zn-rich growth. On the other hand, Co-rich growth leads to a weakly temperature dependent background magnetization that we attribute to Co aggregates. For optimally balanced growth the films are smooth, contain large crystallites ${\sim}200$~nm in lateral size, and exhibit magnetic order well above room temperature, with a Curie point of about 420~K. 

\section{Experimental Methods\label{sec:expt}}

\subsection{Sputtering}
Samples of thin film CoZn were deposited by co-sputtering from elemental targets, with the final film composition controlled by setting the ratio of sputter powers applied to the targets. These films were grown on the ${\sim}100$~nm\ SiO$_{x}$ surface layer of 0.5~mm thick thermally oxidized Si substrates, which were at ambient temperature inside the vacuum chamber. The deposition was done using balanced magnetrons with an argon working gas at a growth pressure of 3.2~mTorr. The growth system used has a base pressure of approximately $1 \times 10^{-8}$~Torr.

\subsection{Annealing} 

\begin{figure}
    \includegraphics[width=8.6cm]{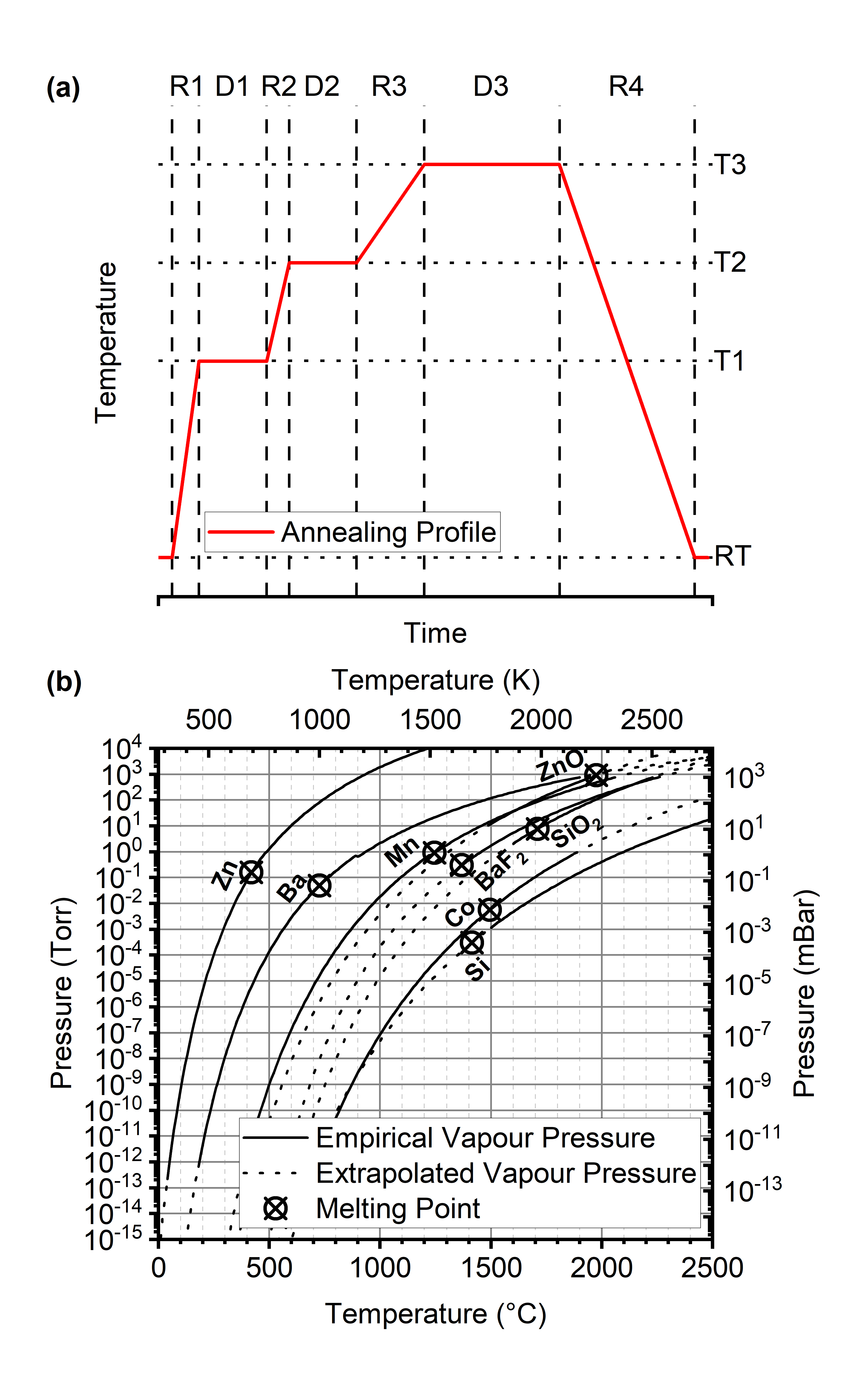}
	\caption{Annealing. (a) General annealing profile schematic. The temperature axis defines temperature set-points ($T1$, $T2$, etc.) above room temperature ($RT$), while the time axis shows representations for the various dwell ($D$) and ramp ($R$) periods. (b) Vapor pressure curves for elements and compounds pertinent to this work. These curves were generated from the Antoine coefficients and the Antoine equation \cite{Haynes2016,Yaws2009}, expressed as $\log_{10}(P)=A-\frac{B}{T+C}$
    , where $P$ is pressure, $A$, $B$, and $C$ are the Antoine coefficients and $T$ is the temperature.\label{fig:AnnealingProfile}}
\end{figure}

The samples were isolated for annealing post-growth by sealing in quartz glass ampules under a vacuum. The tube was evacuated to a roughing pressure of order $10^{-1}\sim 10^{-2}$~Torr before applying heat to the tube with an oxy-acetylene torch until it was fully sealed, then twisted by hand until the ampule containing the samples was sealed and separated from the main tube body. At every step in the process the system is repeatedly pump-purged with 99.9999$\%$ pure argon gas to keep the sample atmospheric environment as free from oxygen as possible.

Annealing was performed using a muffle furnace with a PID temperature controller; temperature measurement was by a K-type thermocouple. The furnace is a single-piece molded ceramic fiber chamber for high thermal efficiency, with heating elements embedded inside the walls of the chamber. Sample temperature stability in the chamber at rated temperature in thermal steady state is within $\pm1$~\degree C, with a temperature uniformity of $\pm10$~\degree C, up to a maximum temperature of 1100~\degree C.

The temperature controller used allows for up to four set-points of ramp rate, temperature, and dwell time to be defined, illustrated schematically in Fig.~\ref{fig:AnnealingProfile}(a). In the case of the samples described here, the typical annealing profile was to use staggered (decreasing) ramp rates ($R1$/$R2$/$R3$) of 5/3/1~\degree C per minute, to temperature setpoints ($T1$/$T2$/$T3$) of 300/350/400~\degree C and a final stage dwell time $D3$ of 48~hours. After annealing, samples were either cooled at the natural rate of the furnace with the heating output turned off, or quenched in water ($R4$). Occasionally, intermittent dwell times ($D1$/$D2$) were used at the 300/350~\degree C temperature setpoints of up to 48~hours. These examples are of the typical process, other values for these parameters were explored in improving the quality (crystalline order and magnetic properties) of films.

\subsection{X-ray diffraction}
\begin{figure}
    \includegraphics[width=8.6cm]{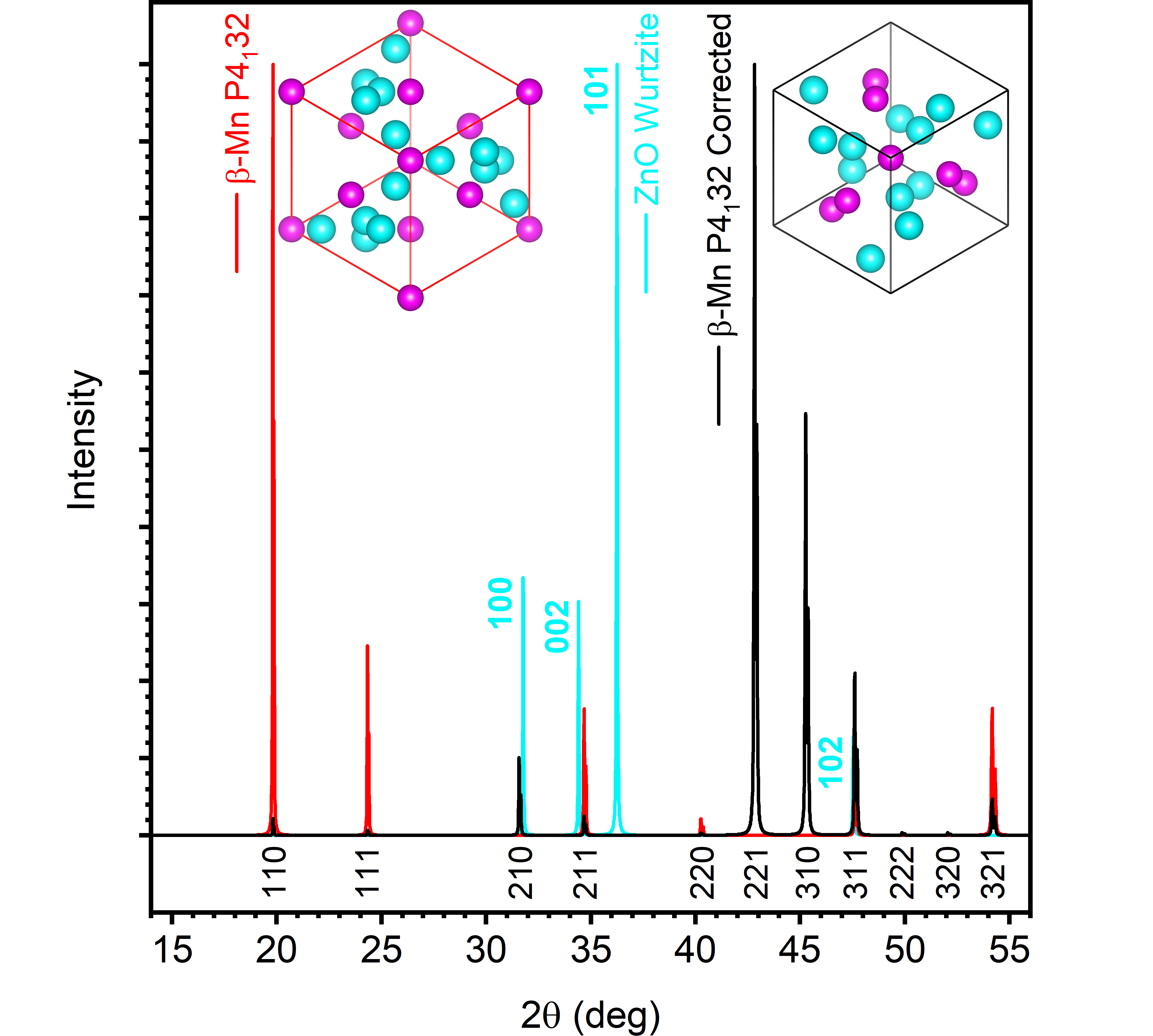}
    \caption{Simulation of the X-ray powder diffraction pattern for the $\beta$-Mn space group P4$_{3}$32/P4$_{1}$32 (space group number 212/213). After accounting for the equivalent isotropic displacement parameters for the 8c and 12d sites \cite{Xie2013c}, the brightest expected reflection shifts from the \{110\} in the basic space group (red), to the \{221\} in the corrected structure (black). The other major component seen in the films herein, ZnO (cyan) \cite{Heller1950}, shows a number of overlapping peaks with the CoZn $\beta$-Mn phase. Structures and simulated data here were produced in VESTA \cite{Momma2011}.\label{fig:powdercell}} 
\end{figure}

X-ray diffraction (XRD) studies were performed using a diffractometer of Bragg-Brentano geometry with a 9~kW rotating Cu anode X-ray source. Monochromatic $K_\alpha$ radiation was achieved using a Ge (220) 2-bounce monochromator, and divergence limited with 5\degree{} receiving Soller slits. 
The expected X-ray diffraction patterns from $\beta$-Mn structure CoZn and ZnO are shown in Fig.~\ref{fig:powdercell}.

\subsection{Electron microscopy and energy-dispersive X-ray spectroscopy analysis}
Imaging of the sample surface and production of lamella cross-section specimens were conducted using a dual-beam focused ion beam/scanning electron microscope (FIB/SEM) instrument. Scanning electron microscopy was performed with an accelerating voltage of 5 kV. Lamella cross-sections were prepared from thin film samples following the standard `lift-out' process \cite{Overwijk1993,Langford2008,Schaffer2012}. The lamellae are estimated to be ${<}100$~nm thick.

Transmission electron microscopy (TEM) and scanning TEM (STEM) were performed on an aberration-corrected microscope with a 300~kV extreme field emission gun source, with the beam monochromated to an energy spread  of ${\sim}0.25$~eV.  High resolution (HRTEM), bright field and high angle annular dark field (HAADF-STEM) imaging modes were used. Energy-dispersive X-ray spectroscopy (EDXS) was performed in the same microscope.

\subsection{Magnetometry}
Magnetometry was performed using a superconducting quantum interference device vibrating sample magnetometer (SQUID-VSM), equipped with a 7~T superconducting solenoid magnet, capable of heating the sample up to 1000~K under a pressure ${<}100$~mTorr. 

Our measurement protocol began with an isothermal hysteresis loop $M(H)$ measurement at room temperature from positive to negative saturating field $\pm H_\mathrm{max} = 8000$~Oe and back to positive before returning to a measurement field of $H_\mathrm{measure} = +50$~Oe. This field was then held constant at this value whilst the temperature was raised to 500~K in order to measure $M(T)$. A further $M(H)$ hysteresis loop was performed at this elevated temperature, which is expected to be above the Curie point $T_\mathrm{C}$ of the CoZn but at which any Co regions will still be strongly ferromagnetic. Fields were applied in the plane of the sample in all cases. 

$M(T)$ curves were fitted from $T = 300$~K to the Curie temperature $T = T_\mathrm{C}$ with an empirical expression that approximates to Bloch's law at low $T$ \cite{Evans2015}, written as:
\begin{equation}
M(T)=M_{300}\times\left(1-\left[\frac{T}{T_{\text{C}}}\right]^{\alpha}\right)^{\beta},
\label{eqn:CurieBloch}
\end{equation}
where $M_{300}$ is the magnetization at room temperature, $\alpha$ is an exponent representing the `Bloch' term describing the behavior in the low temperature region, and $\beta$ is an exponent representing the `Curie' term describing the behavior near the Curie point.

\section{Results\label{sec:results}}

\subsection{Sample growth and X-ray diffraction \label{sec:resultscodepfilms-xrd}}

\begin{figure*}
    \includegraphics[width=17.2cm]{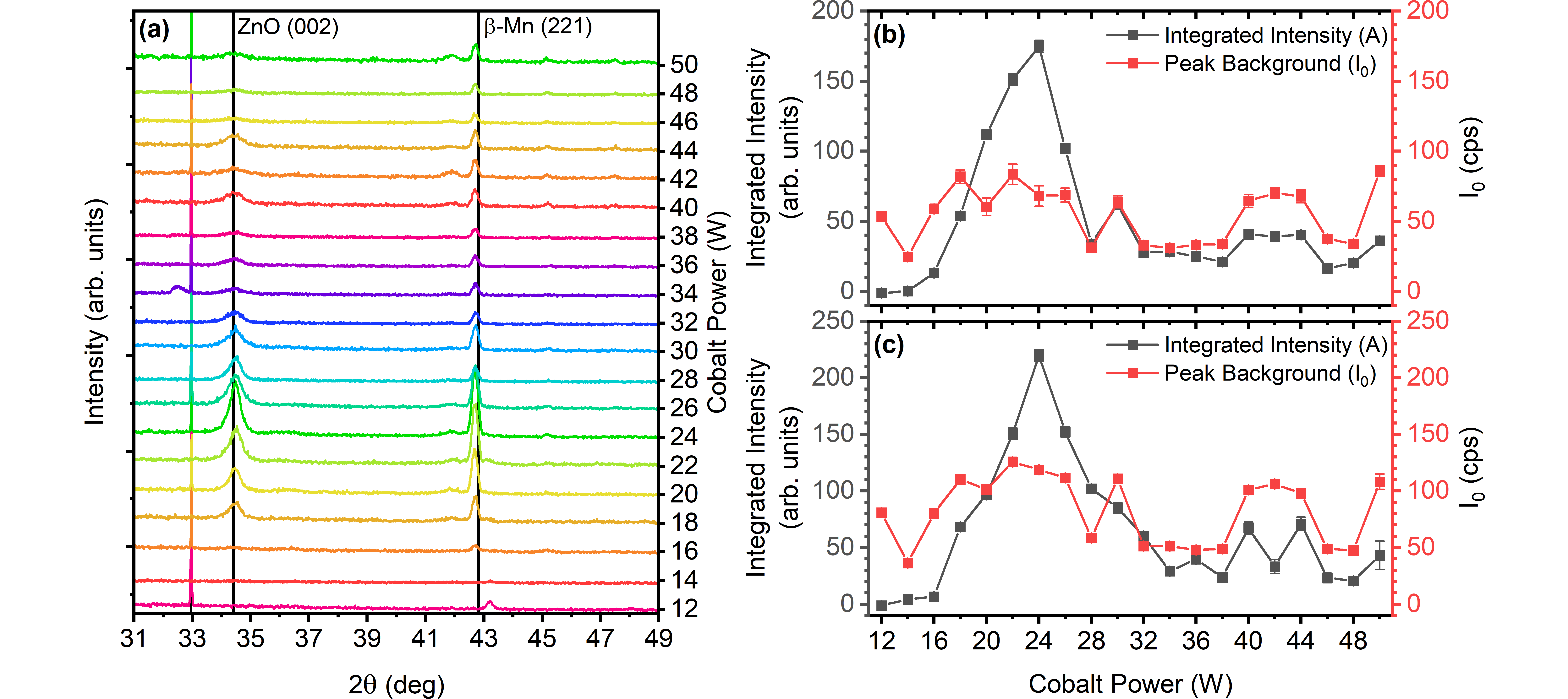}
    \caption{X-ray diffraction analysis of post-annealed sputtered CoZn thin films. (a) XRD spectra for films deposited with various values of Co sputter power (Zn fixed at 25~W). Principal peaks are $(002)$ ZnO at ${\sim}34.45$\degree{} and $(221)$ $\beta$-Mn at ${\sim}42.70$\degree. The expected positions for these are marked by solid black lines. Weaker peaks can be seen at ${\sim}45.2$\degree{} and ${\sim}47.5$\degree{}, which correspond to expected positions for $(310)$ and $(311)$ $\beta$-Mn, respectively. The sharp peak at ${\sim}32.96$\degree{} is the classically `forbidden' $(200)$ peak of Si. Samples here are calibrated to all fall in the range 400\textendash500~\AA\ thick. (b) Summary of Gaussian peak fitting results for the ZnO $(002)$ peak. (c) Summary of Gaussian peak fitting results for the $\beta$-Mn $(221)$ peak.\label{fig:xrd}}
\end{figure*}

X-ray diffraction was used to analyze the crystallographic properties of the films that we grew. To produce samples of varied composition, we co-sputtered films whilst varying the Co power between samples whilst fixing the Zn sputter power at 25~W \footnote{Zn power was chosen to be 25~W based on separate atomic force microscope studies that showed minimum film roughness at this power.}. A post-growth annealing protocol consistent with Fig.~\ref{fig:AnnealingProfile}(a) was developed on the basis of a broader range of XRD studies to optimize the presence of Bragg reflections corresponding to the $\beta$-Mn structure whilst minimizing those corresponding to other phases. 

Our XRD studies have also shown a tendency for zinc to be vaporized from the surface of the films for annealing temperatures over 500~\degree C, resulting in emerging pure cobalt phase Bragg peaks. This is due to the high vapor pressure of zinc at these temperatures (see Fig.~\ref{fig:AnnealingProfile}(b)). 

Following this preliminary work, we acquired XRD spectra for our series of annealed, co-sputtered films, shown in Fig.~\ref{fig:xrd}(a). Comparison with Fig.~\ref{fig:powdercell} shows that most balances of sputter powers, but especially for a Co sputter power around 24~W, yield peaks primarily around 34.45\degree{} and 42.70\degree. These peaks were determined to be the $(002)$ (34.4201\degree) and $(221)$ (42.8237\degree) reflections of ZnO and $\beta$-Mn structure CoZn, respectively, by comparison against all known allotropes and alloy phases of cobalt, zinc and their oxides. 

There are indications of a co-dependence to the appearance and ordering of ZnO with the $\beta$-Mn. Fig.~\ref{fig:xrd}(b) and (c) show the results of fitting the two peaks with a Gaussian according to:
\begin{equation}
I = I_{0}+\frac{A}{w\sqrt{\frac{\pi}{2}}} \exp \left(-2\frac{(2\theta-2\theta_\mathrm{c})^2}{w^2} \right),
\label{eqn:Gauss}
\end{equation}
where $I_0$ is the background intensity, $A$ is integrated intensity under the peak, $w$ is the peak width, and $2\theta_\mathrm{c}$ is the peak center.

It can be seen from Fig.~\ref{fig:xrd}(b) and (c) that the integrated intensity (derived from the fitting of the Gaussian to the peak) of both the ZnO and $\beta$-Mn CoZn rise and fall together. This would indicate some co-dependent behavior even when minimizing the free Zn at higher Co weighting. Furthermore, the $I_0$ term also rises and falls in the same manner which will to some degree be suppressing the integrated intensity under the peak and thus the prominence of the two distinct peaks found in the series. This elevation of the $I_0$ term also indicates that the broadening of the peak becomes non-Gaussian towards the base, which implies that there is a distribution of smaller disordered crystallites. 

Since the $\beta$-Mn (221) peak is strongest for 24~W Co deposition power, we began our investigations in films grown in that way. 

\subsection{Electron microscopy\label{sec:resultscodepfilms-tem}}

\begin{figure*}
    \includegraphics[width=17.2cm]{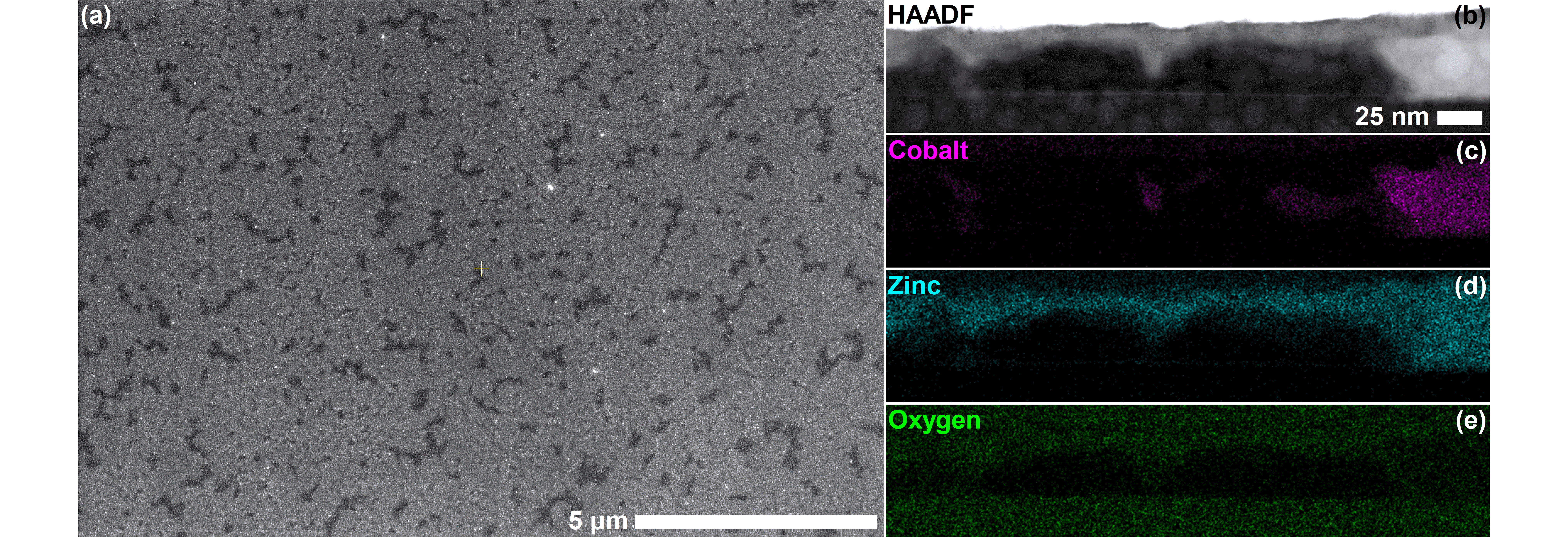}
    \caption{Electron microscopy of an annealed sample of 24:25~W Co:Zn (approximately $460 \pm 40$~\AA{} thick). (a) SEM image taken normal to the sample surface showing contrast differences within the layer. (b) HAADF-STEM image of an extracted lamella cross-section, which reveals the surface contrast to be due to large vacancies. Breakdown of the composite elements expected within the EDXS image around the vacancy defect in panel (b): (c) cobalt signal, (d) zinc signal, (e) oxygen signal. \label{fig:2425}}
\end{figure*}

Fig.~\ref{fig:2425}(a) shows a preliminary SEM image of the surface of the 24:25~W Co:Zn sample taken in the FIB/SEM system prior to laying down the protective coatings needed for preparing the cross-section specimen. This power ratio is the one at which the strongest XRD peak for $\beta$-Mn structure CoZn was observed. The contrast is not uniform, indicating an inhomogeneous surface. Further investigation, by cross-section HAADF-STEM imaging of a lamella taken from a similar region (Fig.~\ref{fig:2425}(b)), shows there to be large vacancies within the film. While the XRD shows that the CoZn material is highly ordered in the $\beta$-Mn structure at this 24:25~W Co:Zn composition, the excess of zinc at high temperatures and low pressures leads to the vaporization of zinc causing these large vacancies to form. The distribution of O, Zn, and Co around these vacancies is shown in the EDXS maps in Fig.~\ref{fig:2425}(c), (d), and (e), respectively. The EDXS maps reveal a ZnO layer that persists as a cap across the vacancy site, while there is almost no cobalt presence within the vacancy except for some aggregates with low/poorly mixed Zn content.

\begin{figure}
	\includegraphics[width=8.6cm]{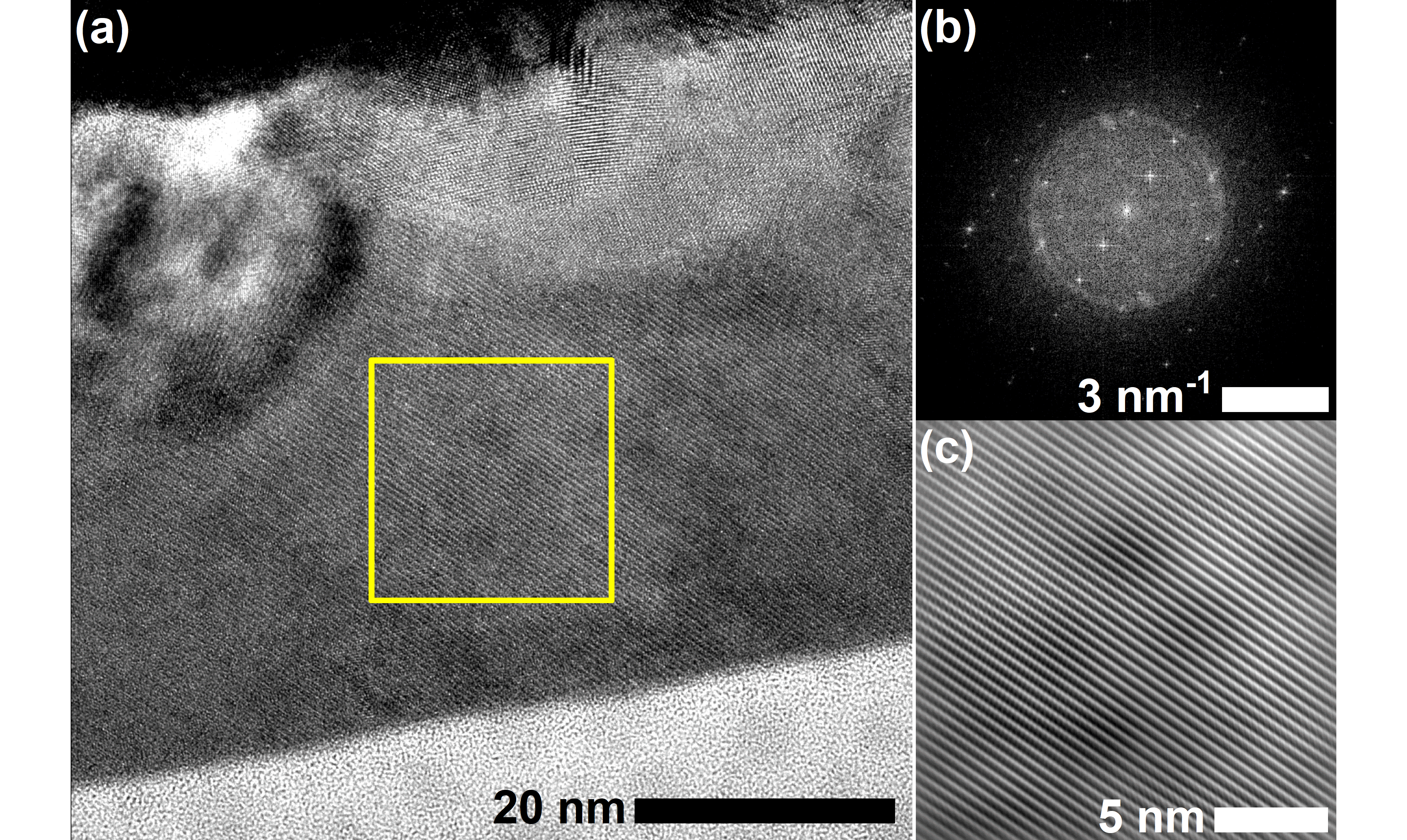}
	\caption{High-resolution TEM of a sample grown with 24:25~W Co:Zn. (a) HRTEM image. Despite the vacancies prevalent throughout the sample, there is extensive crystalline material of 10\textendash120~nm lateral dimensions throughout. (b) A fast-Fourier transform taken centrally from the indicated region in (a); in which the bright spots relate to the crystalline periods present, while the diffuse circle is the disordered/amorphous material (chiefly due to the presence of the SiO$_{x}$ substrate). (c) Inverse fast-Fourier transform image of (b), clearly showing the visible lattice fringes.\label{fig:tem}} 
\end{figure}

High-resolution TEM of the cross-section specimen, shown in Fig.~\ref{fig:tem}, agrees with the XRD in showing a high proportion of crystallographically well-ordered material within the films. From the image contrast in HRTEM Fig.~\ref{fig:tem}(a), we can resolve the crystal structure atop the amorphous SiO$_{x}$ substrate, with the surface of this crystal disrupted by a lower $Z$ (presumably oxide) layer, which has a more granular microstructure. 

Fig.~\ref{fig:tem}(c) shows an image from the central region of the large crystallite in Fig.~\ref{fig:tem}(a); in preparing this image, brightness/contrast and the maximum/minimum thresholds were adjusted in Fourier space in order to better visualize the lattice fringes of the (110) planes in the $\beta$-Mn structure. Fig.~\ref{fig:tem}(b) is the fast Fourier transform of the central region of Fig.~\ref{fig:tem}(a), showing the frequency space response with clear crystalline periodicity overlaid with some contaminant peaks (likely to arise from the ZnO cap) and an amorphous circle from the inclusion of the SiO$_{x}$ substrate.

\begin{figure}
    \includegraphics[width=8.6cm]{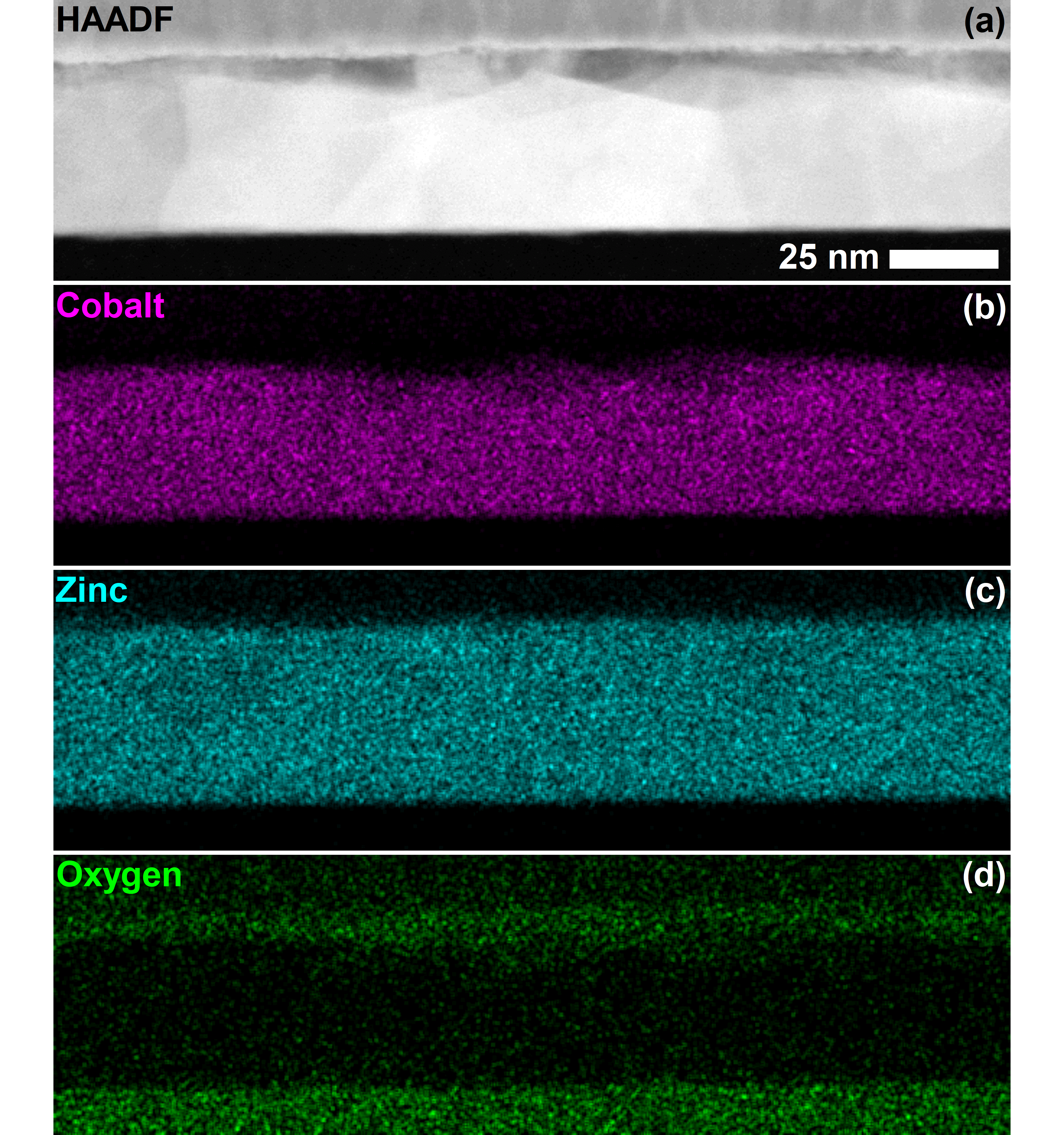}
    \caption{STEM/EDXS analysis of an annealed 36:25~W Co:Zn alloy sample. (a) HAADF-STEM image of EDXS scan region. (b) Cobalt element signal. (c) Zinc element signal. (d) Oxygen element signal. There exists only a small oxide layer atop the film, which is distinct from the Co contribution.\label{fig:edx}}
\end{figure}

Evidently, whilst the XRD pattern from the 24~W Co sample shows strong (221) $\beta$-Mn texture, the TEM reveals large vacancies and a thick ZnO cap blanketed the film, arising from the excessive Zn content. We therefore turn our attention now to a film with a higher Co content. Fig.~\ref{fig:edx} shows TEM analysis of a cross-section cut from an annealed 36:25~W Co:Zn alloy sample. Comparing the high-angle annular dark field (HAADF)-STEM contrast, shown in  Fig.~\ref{fig:edx}(a), with energy dispersive X-ray spectroscopy (EDXS) mapping, displayed in Fig.~\ref{fig:edx}(b\textendash d), we can see that this film consists of CoZn sitting on a substrate with an oxide layer with the excess zinc at the surface. In the case of Fig.~\ref{fig:edx} this oxide layer is much reduced for the 36:25~W Co:Zn over the 24:25~W Co:Zn example in Fig.~\ref{fig:tem}(a) \& \ref{fig:2425}(b\textendash e) due to the correction to the as-deposited element ratios. This reduction in ZnO is consistent with the XRD data presented in Fig.~\ref{fig:xrd}(b).

EDXS results for Co:Zn ratios give a composition of approximately $57 \pm 4$ at.\% Zn to $43 \pm 4$ at.\% Co for 24:25~W, and approximately $54 \pm 3.67$ at.\% Zn to $46 \pm 4$ at.\% Co 36:25~W. These values are within expectations for the stoichiometry of $\beta$-Mn phase CoZn alloy, based on the bulk phase diagram \cite{ASMInternational1992}. 

The example shown in the HAADF-STEM image in Fig.~\ref{fig:edx}(a) contains two fairly significant surface crystallites of ZnO, making tracking of the responsible elements in the corresponding EDXS maps in Fig.~\ref{fig:edx}(b\textendash d) easy. From these maps, one can clearly see that the portion of the sample containing Co is distinct from the oxide portion. However this is not the case for Zn, which overlaps with Co for most of the film thickness but also with O in a thin layer at the top. We can therefore conclude that most of the film is metallic CoZn with a thin ZnO cap. 

\begin{figure*}
    \includegraphics[width=17.2cm]{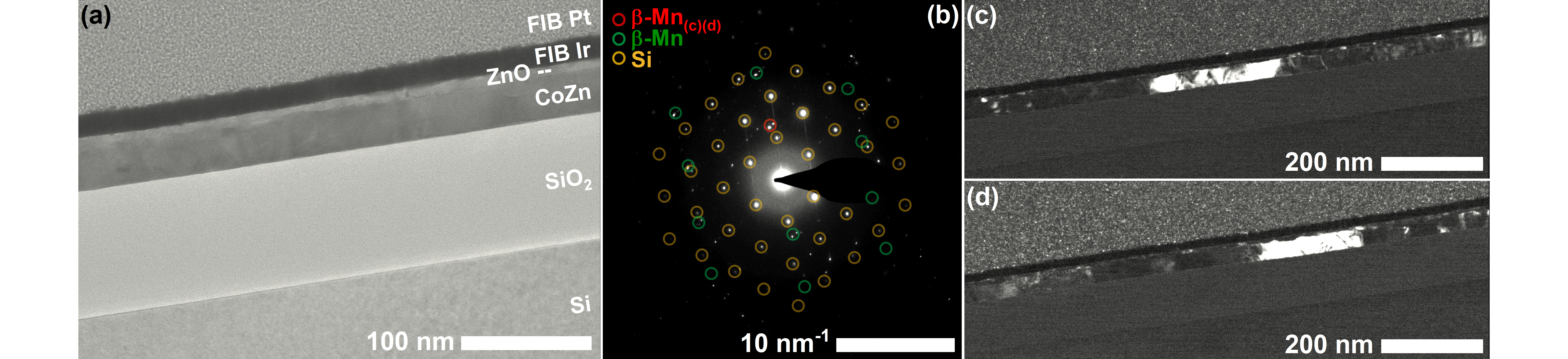}
	\caption{TEM imaging of an annealed sample of 36:25~W Co:Zn. (a) Bright field image with variations in contrast indicative of differences in $Z$-contrast of the different elements involved. (b) Reciprocal space diffraction pattern, showing a combination of Si, SiO$_{x}$, and $\beta$-Mn structure CoZn signals, identified by differently colored circles, from which (c) and (d) were selected. (c,d) Dark field imaging of the 36:25~W Co:Zn alloy sample from an alignment to $\beta$-Mn CoZn peaks in the reciprocal diffraction pattern of (b). These show large plate-like crystallites, over 200~nm in lateral dimension.\label{fig:CoZn3625darkf}}
\end{figure*}

Fig.~\ref{fig:CoZn3625darkf}(a) shows a bright field image of the whole film stack for the 36:25~W Co:Zn sample, which is close to the idealized composition ratio. Panels (b\textendash d) of that figure summarize dark field imaging. Dark field (DF) imaging allows us to image parts of the sample that give rise to specific diffraction peaks within the reciprocal space diffraction pattern, Fig.~\ref{fig:CoZn3625darkf}(b). This results in an orientation specific contrast shift allowing us to view all similarly aligned crystals.

In Fig.~\ref{fig:CoZn3625darkf}(b), which shows a diffraction pattern taken at near-normal incidence to the lamella, we can see from the large number of diffraction spots present that there is a variety of crystal orientations visible. Given the electron beam's spot size, which overlaps the amorphous SiO$_{x}$ and even the Si beneath even that, many diffraction features arise from those regions. However, the contributions from the CoZn film are easily identifiable, and on aligning to them, DF imaging results in several large crystallites being readily apparent. Both of the examples given in Fig.~\ref{fig:CoZn3625darkf}(c) and (d) are over 200~nm in lateral dimension, and have minimal zinc oxide contaminating the surface.

\subsection{Magnetic properties\label{sec:magprop}}

\begin{figure*}
    \includegraphics[width=17.2cm]{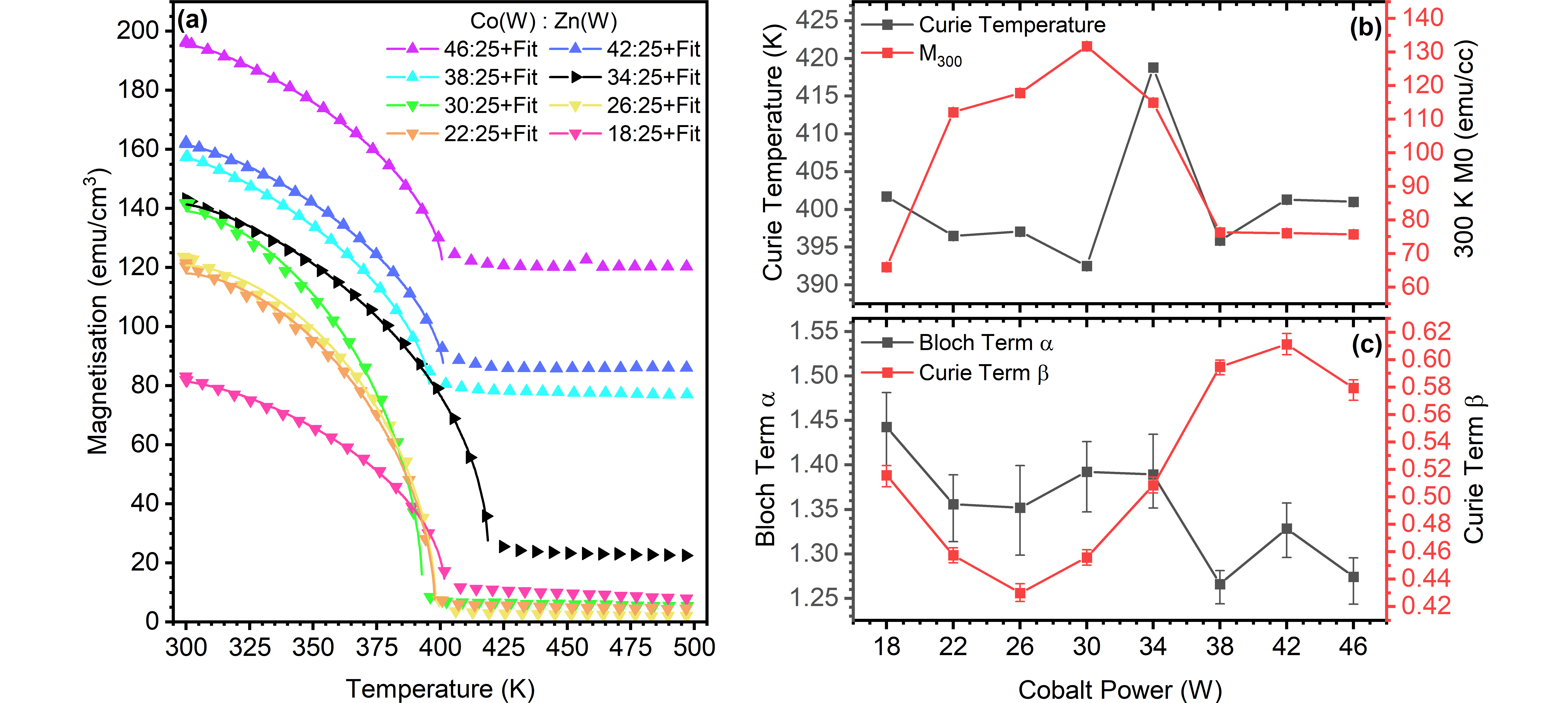}
    \caption{Temperature dependent magnetic measurements of annealed sputtered CoZn thin films. (a) $M(T)$ results for the CoZn films deposited at growth power ratio 18\textendash46:25~W Co:Zn. Data points represent experimental results, the solid line shows the result of fitting, as described in the text. The extent of the solid line shows the region that was fitted.  (b) $T_\mathrm{C}$ and $M_{300}$ versus composition. (c) Empirical Curie-Bloch fitting exponents $\alpha$ and $\beta$ versus composition.\label{fig:MTComp}}
\end{figure*}

The magnetic properties of our films were measured by SQUID-VSM. All of our CoZn films display $T_\mathrm{C}$ above room temperature, as shown in the magnetization versus temperature curves of Fig.~\ref{fig:MTComp}(a). In most cases, the value of $T_\mathrm{C}$ lies between 390\textendash405~K. However, around the idealized composition ratio (of 34:25 W Co:Zn as determined by XRD shown in Fig.~\ref{fig:xrd} and STEM/EDXS shown in Fig.~\ref{fig:edx}) there is a significant 
jump to a higher $T_\mathrm{C}$ of approximately 420~K, shown in Fig.~\ref{fig:MTComp}(a) and summarized in Fig.~\ref{fig:MTComp}(b) where we plot $T_\mathrm{C}$ as a function of Co deposition power. This value is comparable to bulk values \cite{Xie2013c,Tokunaga2015,Bocarsly2019}.

Further examination of Fig.~\ref{fig:MTComp}(a) shows that the magnetization retains a substantial value above $T_\mathrm{C}$ for Co content higher than this ideal value. Even at the ideal composition there remains some retained background magnetism above $T_\mathrm{C}$. This is likely to arise from the development of cobalt-rich aggregates above the ideal power ratio of 34:25~W Co:Zn, which remain ferromagnetic to much higher temperatures: the bulk $T_\mathrm{C}$ of Co is ${\sim}1400$~K.

These $M(T)$ curves were fitted with the empirical approximation to Bloch’s law for the ferromagnetic phase given by Eq.~\ref{eqn:CurieBloch}. All curves had a linear background subtraction applied based on the above-$T_\mathrm{C}$ magnetization prior to fitting. Hence we define $M_{300}$ as the magnetization of the CoZn film, i.e. above the background level, at 300~K. 

Fitting of the M(T) curve shows a few interesting features beyond the spike seen in $T_\mathrm{C}$ around the ideal composition ratio. Fig.~\ref{fig:MTComp}(b) also shows the room temperature magnetization $M_{300}$ to have a region of elevated values ${\sim}120$~emu/cm$^3$ specifically between our case of strong crystallinity in XRD at 18:25~W Co:Zn to just beyond the compositional ideal seen around 34\textendash36:25~W Co:Zn. Beyond this limit the room-temperature magnetization drops to approximately 75~emu/cm$^3$. We interpret this as meaning that the remainder of the film can become somewhat depleted in Co as the Co-rich aggregates form. 

In Fig.~\ref{fig:MTComp}(c) we show the Bloch and Curie exponents $\alpha$ and $\beta$. In a simple ferromagnet, $\alpha = \frac{3}{2}$ according to the Bloch law, which is valid at temperature well below $T_\mathrm{C}$. We do not have data here in that regime, but nevertheless, the values of $\alpha$ that were returned by our fit are not far below this value. 

Meanwhile, $\beta$ is a critical exponent that has a value $\frac{1}{2}$ according to a simple Landau model of the ferromagnetic phase transition. We find values close to this, especially near the `ideal' composition of 34:25~W Co:Zn.  

\section{Conclusion\label{sec:conclusion}}

To summarize the results, the XRD shows that co-deposited films of CoZn annealed up to 400~\degree C have a stable state of $\beta$-Mn phase over a wide range of sputtered composition ratios. The resulting phase prevails due to the high mobility of zinc at low temperatures. The limiting factors appear to be cases where the composition is exceptionally high in one or other of the two elements. In the case of high zinc content we are limited by the vapor pressure and annealing temperature of the films leading to a substantial ZnO cap; and in the case of high cobalt content we are limited by the accrual of cobalt-rich aggregates. 

While $\sim$24~W deposition power for cobalt shows the strongest $\beta$-Mn crystalline response by XRD, it was observed by electron microscopy that these films were full of vacancies, as well as possessing the thick ZnO cap. In fact $\sim$36~W deposition power for cobalt led to a better volume fraction of $\beta$-Mn CoZn with a much more uniform film.

Crystallinity within the films appears limited, with grains not observed to grow above ${\sim}200$~nm in-plane, regardless of annealing time. The mechanisms responsible for this are unknown, but it is proposed that due to zinc having a high affinity for oxygen binding and alloy formation that contamination from the vacuum could be an issue. Alternatively, this could be a mechanical restriction based on the diffusion of the elements close to stoichiometry, or limited by energetic requirements to rearrange masses of metastable crystallites.

Annealed films of co-sputtered CoZn of $\beta$-Mn phase are ferromagnetic at and above above room temperature. 
Around the `best' composition by volume ratio (36~W Co deposition power), a higher T$_{\textrm{C}}$ than the rest of the series that show good $\beta$-Mn crystalline structure is obtained. For higher Co deposition powers we observed a weakly temperature dependent background magnetization that we attribute to excess Co precipitating out into aggregates. 

Both the room-temperature magnetization and the $T_{\textrm{C}}$ of thin-films close to the ideal composition of Co:Zn to give a full volume fraction of $\beta$-Mn compare well with the bulk crystal literature results of Tokunaga \emph{et al.} (2015) \cite{Tokunaga2015}. The positive comparison of these features, and the expectation that the magnetic phase diagram would follow the template of the B20 materials \cite{Yu2010,Muhlbauer2011b}, leads to the potential for these films to also support chiral magnetic textures. Our results suggest that attempts at epitaxial growth and to seek chiral magnetism are fruitful future directions for research.

The data associated with this paper are openly available from the University of Leeds Data Repository at \cite{data}.


\begin{acknowledgments}
We would like to thank Dr Gavin Stenning and Dr Daniel Nye for help on the Rigaku Smartlab and Quantum Design MPMS3 SQUID-VSM instruments in the Materials Characterisation Laboratory at the ISIS Neutron and Muon Source.

We would like to thank Mr John Harrington and Dr Zabeada Aslam for help on the Helios G4 DualBeam FIB and Titan$^3$ electron microscope instruments at the Leeds Electron Microscopy and Spectroscopy Centre.

This work was supported in part by the ISIS Facility Development Studentship scheme.

\end{acknowledgments}

\bibliography{library_PostVIVA}

\end{document}